\documentclass[useAMS,usenatbib]{mn2e}

\usepackage{bbm}
\usepackage{mathrsfs}

\newcommand\de{\delta}

\newcommand\ze{\zeta}

\renewcommand\th{\theta}

\newcommand\rh{\rho}

\newcommand\pa{\partial}

\newcommand\eg{\emph{e.g.}}

\newcommand\beq{\begin{equation}}
\newcommand\eeq{\end{equation}}
\newcommand\bea{\begin{eqnarray}}
\newcommand\eea{\end{eqnarray}}
\newcommand\bal{\begin{align}}
\newcommand\eal{\end{align}}

\newcommand\fr{\frac}

\newcommand\rms[1]{_{\mathrm{#1}}}

\newcommand\half{{\textstyle \frac{1}{2}}}
\newcommand\ap{\approx}

\renewcommand\bal{\mbox{\boldmath$\alpha$}}

\newcommand\cH{\mathcal{H}}
\newcommand\cHi{\mathcal{H}\rms{i}}
\newcommand\cP{\mathcal{P}}
\newcommand\ai{a\rms{i}}

\newcommand\rhr{\rh\rms{r}}
\newcommand\rhv{\rh\rms{v}}
\newcommand\Ti{T\rms{i}}
\newcommand\Tph{T\rms{phys}}

\newcommand\pt{\propto}
\newcommand\Heq{H_{\rm eq}}
\newcommand\aeq{a_{\rm eq}}

\title[Does inflation erase curvature and relics?]{Has inflation really solved the problems of flatness and absence of relics?}
\author[Richard  Lieu]{Richard Lieu\thanks{E-mail:
lieur@uah.edu}\\ Department of Physics, University of Alabama,
Huntsville, AL 35899, U.S.A.\\}

\begin{document}

\maketitle

\label{firstpage}

\begin{abstract}
Among the three cosmological enigma solved by the theory of inflation, {\it viz.}~(a) large scale flatness, (b) absence of monopoles and strings, and (c) structure formation, the first two are addressed from the viewpoint of the observed scales having originated from very small ones, on which the density fluctuations of the curvaton and relics are {\it inevitably} of order unity or larger.
By analyzing strictly classically (and in two different gauges to ensure consistency) the density evolution of the smoothest possible pre-inflationary component -- thermal radiation -- it is found that the O(1) statistical fluctuations on the thermal wavelength scale present formidable obstacles to the linear theory of amplitude growth by the end of inflation.   Since this wavelength scale exited the horizon at an early stage of inflation, it severely limits the number of e-folds of perturbative inflation.  With more e-folds than $\approx 60$ there will be even larger fluctuations in the radiation density that ensures inflation keeps making `false starts'.  The only `way out' is to invoke a super-homogeneous pre-inflationary fluid, at least on small scales, adding to the fine-tuning and preventing one from claiming that inflation simply `redshifts away' all the relic inhomogeneities; {\it i.e.} the theory actually provided no explanation of (a) or (b), merely a tautology.

\end{abstract}


\noindent\noindent

\section{Introduction}

The theory of inflation (\citep{gut81,alb82,lin82}) is thought to have provided attractive solutions to a number of cosmological problems, including the large-scale homogeneity and flatness of the universe.  It garnered strong support from the \texttt{COBE} and \texttt{WMAP} observations of the cosmic microwave background radiation (CMBR) (\citep{smo92,ben03,spe07,hin09,kom11}).  Quantum fluctuations in a scalar inflaton field can explain the origin and near scale-invariant spectrum of the primordial density perturbations, although getting the amplitude right requires fine tuning.

Apart from the central question of structure formation, however, we wish to revisit some of the other key issues of cosmology that inflation is claimed to have satisfactorily addressed, as they too lend as crucial evidence in favor of the theory if the logic is sound.  In particular, the reason for the `global' flatness of space and the absence of strings and monopoles is, in the context of inflation, due to the ability of the rapid expansion in drastically diluting these features.  This means, at the onset of inflation the density of curvatons and relics was probably on par with that of the inflaton.  On the relevant cosmological scales that were deeply embedded within the horizon at the time, the {\it fluctuations} in the densities of the unwanted components are then initially very large, and are likely to be governed by quantum effects.  In general, it is reasonable to assume that the smoothest component is the one with the largest number density, {\it viz.} the massless particles.  Therefore in this paper we examine closely the state of a pre-existing photon population on small scales, by taking the universe before inflation as radiation dominated.  This will set lower limits on the severity of the initial fluctuations.

Indeed, early versions of inflation theory did envisage a pre-inflationary Friedmann-Robertson-Walker (FRW) stage.  One of the claimed advantages of the theory was that inflation effectively erased all traces of that `pre-historic' phase, although it was realized early (\citet{fri84}) that this is not strictly true; what it does is not to eliminate perturbations but to stretch them to unobservable scales.  This leaves open the question of whether there could be perturbations on very tiny scales that are stretched to observable size.  It has been shown (\citet{mag07}) that without drastic modifications such perturbations could not explain the power spectrum of perturbations.  Here we wish to argue that under rather general conditions these perturbations would both be inevitable and far too large to be consistent with either perturbation theory or observation.  This is because thermal radiation has O(1) statistical fluctuations on scales comparable to the thermal wavelength, and such scales have been stretched to cosmologically relevant ones or beyond.

In yet another manner of speaking, while it has previously been argued (\citet{vac98}) that inflation requires sufficiently fine-tuned initial conditions to ensure ultra-smoothness in the relevant part of the universe at the onset of the process,  we suggest that it is essentially impossible to satisfy the condition if there is a pre-inflationary FRW stage consisting primarily of radiation or other even less homogeneous components.  The only way forward is to postulate a universe that was already extremely smooth before inflation, {\it i.e.} the theory does not explain flatness and the absence of relics, rather takes them for granted.


\section{The pre-inflationary thermal phase}

To focus the readers' attention upon the effect we wish to address, we shall here-and-after consider inflation driven by a scalar field of small density fluctuations $\de\rho_v$ relative to the radiation $\de\rho_r$, and with an equation of state close to $\rho_v = -p_v$. Our conclusion is insensitive to such assumptions.

It is further assumed that by the onset of inflation the patch of the universe from which today's Hubble volume evolved contained a fluid of massless (or light) scalar particles, termed `radiation' here for brevity, with $p_r = \rho_r/3$ (or $w=1/3$) and interactions weak enough to treat them as free, but strong enough to maintain thermal equilibrium.  Moreover, like \citet{bis13} our attention is restricted to {\it adiabatic} perturbations that exit the horizon, even though in this two-fluid model there can be isocurvature ones. This paragraph of our {\it ansatz} is important and requires scrutiny, as there are three aspects to it.

First, we shall only be concerned with (a) the {\it statistical} fluctuations of the thermal phase and not (b) the quantum fluctuations of the fluid.  The distinction was pointed out by \citet{bis13}, who also showed in their Appendix B that so long as the fluid particles are light and the temperature $T$ is much below the Planck mass $m_P$ the fluctuations are primarily statistical in nature anyway.  If the particles are massive there will be the {\it additional} effect of (b), but as we shall show, (a) by itself is in general already significant enough to present formidable challenges to standard inflation theory, hence the inclusion of (b) will not alter the conclusion.

Second is the question of thermalization of a fluid in the pre-inflationary universe, which entails kinetic and chemical equilibrium, attainable when the dominant scattering exceeds the Hubble parameter.  Generic arguments point to ineffective interactions at $T > \alpha^2 m_P \ap 10^{16}$~GeV where $\alpha$ is the fine structure constant (\citet{kol90}, section 3.5), although GUT inflation at $T \sim 10^{15}$~GeV lies outside this range.  In the more recent literature on this topic, {\it viz.}~\citet{all10},~\citet{maz11}, and \citet{bas13}, the first two are written in the context of the reheating epoch, while the last the inflationary epoch itself which relates more closely to our current work.  According to \citet{bas13} the light particles thermalize most readily, which lends support to our {\it ansatz}.  In any case, equilibrium or lack thereof, it shall be shown towards the end of our paper that unless the power spectrum of fluctuations at the onset of inflation departs {\it drastically} from its thermal value on scales smaller than the thermal wavelength (or more fundamentally the average inter-particle spacing, the scale where the relative density fluctuations usually reaches unity) in the direction of a sharp cutoff, inflation cannot proceed in the manner intended to explain observations.  Since this critical scale is expected to lie well within the inflation horizon, thermalization on the relevant scales should proceed much more easily -- there is certainly little justification for the said cutoff.

Third, as each mode of statistical fluctuation exits the horizon there are certainly `adiabatic' solutions that describe its evolution.  As mentioned, the two-fluid system of radiation and inflaton also supports isocurvature perturbations. We choose here not to consider them, because the two kinds of perturbations are orthogonal to each other and evolve independently, {\it i.e.} our contention of raising a flag about the already large adiabatic fluctuations that invalidate the standard approach to the whole problem will not be affected by the inclusion of isocurvature effects.


Turning to the rest of the model, the background space-time is taken to be homogeneous, isotropic and spatially flat.  We consider only scalar metric perturbations, first in the longitudinal or Newtonian gauge, then in the synchronous gauge, then cross compare the results to find consistency.  In the former, we assume that the anisotropic stress is negligible, {\it i.e.}~the two invariant potentials are equal: $\Phi=\Psi$.  Thus we can write the perturbed metric as \beq ds^2 = a^2(\tau)[(1+2\Phi)d\tau^2 -(1-2\Phi)dx^idx^i], \label{newt}  \eeq where $\tau$ is the conformal time ($ad\tau=dt$).  In the latter the metric is \beq ds^2 = a^2(\tau)[d\tau^2-(\de_{ij} + h_{ij}) dx^i dx^j], \label{sync}  \eeq where \beq h_{ij} = -\frac{1}{3}h\de_{ij} -6(\partial_i \partial_j -\frac{1}{3} \de_{ij} \nabla^2)\eta \label{h}  \eeq are respectively the trace and traceless contributions to $h_{ij}$ (\citet{ma95}).

We suppose that the vacuum energy starts to dominate at an initial time $\tau_i$ at which $\rhr=\rhv$, where $\rhr$ stands for the energy density of the radiation and $\rhv$ for that of the inflaton field.  Inflation ends at a reheating time $\tau_r$ when the vacuum energy is converted to radiation.  Since $\rhv$ is nearly constant during the inflationary era, this means that the physical temperature $\Tph$ of the radiation is approximately the same at $\tau_i$ as the reheating temperature just after $\tau_r$.  This is a large temperature, but still some orders of magnitude below the Planck energy.  It is convenient to use a `comoving temperature' $T=a\Tph$ which is nearly constant outside the inflationary period.  More precisely, $g^{1/3}T$ is constant, where $g$ is the number of helicity states of massless particles (with fermions counted as $\fr{7}{8}$); for the universe since reheating, allowing for the effect of neutrino decoupling and electron-positron annihilation,
 \beq g^{1/3}T = \left(\fr{43}{11}\right)^{1/3}T_0, \eeq
where $T_0=2.7$K is the CMB temperature today.  Thus the approximate equality of the physical temperatures before and after inflation means that
 \beq \fr{T_i}{T_{\rm rh}} =\fr{T(\tau_i)}{T(\tau_r)}\approx \fr{a_i}{a_r}
= \fr{a(\tau_i)}{a(\tau_r)}=e^{-N}, \eeq
where $N$ is the number of $e$-folds of inflation.  Equivalently, the comoving temperature in the pre-inflationary phase is
 \beq \Ti \approx 0.3 e^{-N}T_0. \label{Ti}  \eeq

It is now necessary to be precise about the power spectrum of the radiation, which we remind the reader consists of thermal light scalar particles.  On scales above one thermal wavelength, {\it i.e.}~$k<T$ it is given by standard thermodynamics as \beq |\de_{\rm th}|^2 = \cP(k)=\fr{k^3}{2\pi^2}P(k)
 = \fr{60 k^3}{\pi^4 T^3} \qquad (k\ll T). \label{PSmallk}  \eeq
In particular, $|\de_{\rm th}|^2$ is O(1) on the wavelength scale $k \approx T$ (note also that since both $k$ and $T$ are `comoving', $k/T=k\rms{phys}/T\rms{phys}$).  This is important because $k \approx T$ length scales exited the horizon at an early stage during inflation.  The situation for the even smaller scales of $k>T$ that exited the horizon later is more complicated and requires the separate treatment presented in \citet{lie13} where it is shown that \beq \cP(k)=\fr{10 k^4}{\pi^4 T^4}
 \qquad (k\gg T). \label{DeLargek}  \eeq  As will be argued below,
the observable cosmological scales today would map back to the $k \gg T$ ones unless inflation has a marginal number of e-folds.

\section{Evolution equations: the longitudinal gauge}

We turn to the question of how the thermal radiation perturbations on the scale of one thermal wavelength or larger, (\ref{PSmallk}), will evolve through the inflationary era.

To begin with, the conformal Hubble parameter $\cH=\dot a/a=aH$ where $\dot a\equiv da/d\tau$ is given by
 \beq \cH^2=\fr{8\pi G}{3}a^2\rho,\eeq
where $\rho=\rhr+\rhv$ is the total density.  Of course, $\rhr\pt a^{-4}$ while in the early stages of inflation we may assume that $\rhv$ is a constant.  It is then convenient to introduce the dimensionless variable \beq y=\fr{a}{\ai} \label{ydef} \eeq  so that
 \beq \cH^2=\fr{\cHi^2}{2}\left(y^2+\fr{1}{y^2}\right).
 \label{H2}  \eeq
(This equation would \emph{not} hold in the late stages of inflation, when the scalar inflaton field $\phi$ starts to roll down towards its minimum, $\dot\phi^2$ is no longer negligible compared to the potential energy $V(\phi)$, so the `vacuum' energy becomes inhomogeneous and $c\rms{s}^2=\dot P/\dot\rho$ starts to deviate from $1/3$, indeed becoming large and negative.  For the moment we exclude that era; we will not need explicit solutions there, because we can use a conservation law.)

Now the gauge-invariant potential $\Phi$ satisfies a second-order evolution equation (see for example \citet{pet09}),
 \beq \ddot\Phi+3(1+c\rms{s}^2)\cH\dot\Phi+
 [2\dot\cH+\cH^2(1+3c\rms{s}^2)]\Phi-c\rms{s}^2\nabla^2\Phi=0. \label{evol0} \eeq
If we are interested in a universe that contains only radiation and vacuum energy, in which the density fluctuations in the latter may be neglected as part of our {\it ansatz} (see the end of section 1) we may set $c_s^2=1/3$, obtaining
 \beq \ddot\Phi+4\cH\dot\Phi+
 2(\dot\cH+\cH^2)\Phi+\frac{1}{3}k^2\Phi=0. \label{evol1}  \eeq
This equation too will \emph{not} hold in the later stages of inflation.  Denoting derivatives wrt $y$ by primes, we have for any function $f$, $\dot f=\cH yf'$.  Thus (\ref{evol1}) becomes
 \beq \cH^2y^2\Phi''+(5\cH^2y+\cH\cH'y^2)\Phi'
 +2(\cH^2+\cH\cH'y)\Phi+\frac{k^2\Phi}{3}=0. \label{evol2}  \eeq
From (\ref{H2}) we also have
 \beq \cH\cH'y= \fr{\cH\rms{i}^2}{2}\left(y^2-\fr{1}{y^2}\right).
 \label{H'}  \eeq
Substituting (\ref{H2}) and (\ref{H'}) into (\ref{evol2}) and rearranging the terms gives
 \beq y^2(y^2\Phi''+6y\Phi'+4\Phi)+y^{-2}(\Phi''+4y\Phi')+q^2\Phi=0,
 \label{evol3}  \eeq
where we have written
 \beq q^2=\fr{2}{3}\fr{k^2}{\cHi^2} =\fr{1}{3}\fr{k^2}{H^2 a^2}. \label{qdef} \eeq
(Note that $\cHi=\ai H\rms{i}$, but $H\rms{i}$ is \emph{not} the near-constant Hubble parameter $H\approx H\rms{infl}$ during inflation; at $\tau_i$, $\rho=2\rhv$ and therefore $H\rms{i}=\sqrt2 H\rms{infl} \approx \sqrt{2} H$).

In (\ref{evol3}) we have chosen to group the terms in such a way that if $\Phi\sim y^n$ they behave respectively like $y^{n+2}$, $y^{n-2}$ and $q^2y^n$.  The point at which $\rho\rms{r}=\rho\rms{v}$ is given by $y=1$.  For $y\ll 1$, we are deep in the radiation-dominated regime and the first bracket is small; while for $y\gg 1$ vacuum energy dominates, and the second bracket is small.  The time of horizon exit is defined by $k^2/\cH\rms{ex}^2=1$.  Since this occurs during inflation, and well after radiation-vacuum equality, we must have $q\gg 1$, and then $y\rms{ex}^2=3q^2$.

\section{Evolution before and during inflation}

Let us first consider the solution well inside the radiation-dominated era, when $y\ll 1$.  Then the first bracket in (\ref{evol3}) is negligible and it is straightforward to solve the equation.  The known solution is
 \beq \Phi = \fr{A}{2q^2y^2}\left(\fr{\sin qy}{qy}-\cos qy\right) +\fr{B}{2q^2y^2}\left(\fr{\cos qy}{qy}+\sin qy\right),
 \label{Ph_r}  \eeq
where $A$ and $B$ are arbitrary constants.  If we impose the natural condition that the perturbation remains finite at early times, we require $B=0$.

The density contrast is given in terms of $\Phi$ by
 \beq \cH(\dot\Phi+\cH\Phi)+\frac{1}{3}k^2\Phi
 =-\half\cH^2\de, \label{Pois}  \eeq
or
 \beq \de =-2[y\Phi'+\Phi+q^2(y^2+y^{-2})^{-1}\Phi].
 \label{de_Ph}  \eeq
Over most of the range, where $1/q\ll y\ll 1$, we can ignore terms involving inverse powers of $qy$, so (\ref{Ph_r}) and (\ref{de_Ph}) give
 \beq \de_r \approx A\cos qy, \label{Acoeff} \eeq
which represents the expected oscillation with approximately constant amplitude.  The typical value of $A$ is given by the expected amplitude of thermal fluctuations.

Now let us turn to the other extreme, when $y\gg 1$ and vacuum energy is dominant.  In that case, the second bracket in (\ref{Ph_r}) is negligible.  It is again possible to find an explicit solution to the equation, by changing variable to $1/y$.  The solution in this case is very similar:
 \bea \Phi &=& \fr{C}{2q^2y^2}\left(\fr{\sin q/y}{q/y}-\cos q/y\right) + \nonumber \\
 &&\ + \fr{D}{2q^2y^2}\left(\fr{\cos q/y}{q/y}+\sin q/y\right),
 \label{Ph_v}  \eea
where again $C$ and $D$ are arbitrary constants.

We still need to connect these two solutions (\ref{Ph_r}) and (\ref{Ph_v}), to relate the constants $C$ and $D$ to $A$.  In the intervening region, near the crossover point $y=1$, it is not possible to find an explicit solution.  But we can get a good idea of the relation by simply assuming that (\ref{Ph_r}) applies throughout the whole of the radiation-dominated era, $y<1$, and (\ref{Ph_v}) throughout $y>1$, and matching the two at $y=1$.  Assuming that $q$ is large enough for the first terms in all the brackets to be negligible, the boundary conditions on $\Phi$ and $\Phi'$ give us
 \bea A\cos q&=&C\cos q-D\sin q,\nonumber\\
 A\sin q&=&-C\sin q-D\cos q, \nonumber\eea
yielding the solution
 \beq C=A\cos 2q,\qquad D=-A\sin 2q. \eeq
These relations might be modified by a more accurate treatment, but $C$ and $D$ would still be related to $A$ and $B$ by a symbolic rotation; the main effect would be to change the effective rotation angle from $2q$.  Note also that even if we did not impose the initial condition of $B=0$, but instead took it be comparable to $A$, $C$ and $D$ would still be of a similar order of magnitude to $A$.  From the remarks made after (\ref{Acoeff}) therefore, one expects \beq |C| \approx |D| \approx |\de_{\rm th}|, \label{CD} \eeq where $|\de_{\rm th}|$ denotes the thermal oscillation amplitude at the scale of interest.

The behavior of the two terms in (\ref{Ph_v}) when $y>q$ and the perturbation is outside the horizon is very different.  The leading terms give
 \beq \Phi\approx \fr{C}{6y^4}+
 \fr{D}{2}\left(\fr{1}{q^3y}+\fr{1}{2qy^3}\right).
 \label{bigy}  \eeq
The corresponding terms in $\de$ are given by (\ref{de_Ph}), and are
 \beq \de\approx \fr{C}{y^4}-\fr{Dq}{y^5}. \eeq
It is interesting that the terms in $y^{-1}$ and $y^{-3}$ cancel.  Thus although the dominant term in $\Phi$ is the $D$ term, in $\de$ it is the $C$ term.  Equivalently the two solutions are, in terms of the radiation density contrast and peculiar velocity divergence $\de_r$ and $\th_r=i{\bf k} \cdot {\bf v}_r$, \bea &\de\rms{r}& = C,~\th\rms{r} = -\fr{k^2}{4Ha^2}C =-\fr{3q^2 H}{4}C;~{\rm or}~ \nonumber\\
&\de\rms{r}& = -\fr{qD}{y} = -\fr{k}{\sqrt{3}Ha}D,~\th\rms{r}=-\fr{\sqrt{3}k}{4a}D, \label{cd} \eea where $\th_r$ is a special case of the generic equation for the total divergence $\th$: \beq k^2 (\dot\Phi + H\Phi) = (\cH^2-\cH a) \th = 4\pi G a^2 (\rho+p) \th \label{th_L} \eeq in the limit of $p_v=-\rho_v$ and $p_r = \rho_r/3$ when $(\rho + p) \th = \sum_i (\rho_i + p_i)\th_i = (\rho_r + p_r)\th_r = 4\rho_r \th_r/3$.

\section{Reheating and subsequent evolution}

It is useful to define the quantity $\ze$, known as the total curvature perturbation, that has showed itself to be particularly useful after a mode exited the horizon:
 \beq \ze = \Phi-\fr{\de}{3(1+w)} = \Phi+\fr{\cH(\dot\Phi+\cH\Phi)+\frac{1}{3}k^2\Phi}
 {\cH^2-\dot\cH}, \label{zedef}  \eeq
Note that $\ze$ is infinite {\it only} in the case of single field inflation of finite $\de_v$ but vanishing $1+w_v$, {\it i.e.}~$\ze$ is finite once there are other, $1+w_i > 0$ components present.
In terms of the variable $y$, (\ref{zedef}) can be written as
 \beq \ze = \Phi+\fr{y^4}{2}
 \left(y\Phi'+\Phi+\fr{q^2}{y^2}\Phi\right). \eeq
Using the evolution equation, one can then show that
 \beq \dot\ze=
 -\fr{k^2}{3}\fr{\cH\Phi+\dot\Phi}{\cH^2-\dot\cH}.
 \label{zedot} \eeq
Both the numerator and denominator of $\ze$ are small, but the quotient need not be, as we shall find out.


Let us evaluate these quantities for the solutions given above in the inflationary era.  We are interested in the values reached well outside the horizon, so we may use the late-time expression (\ref{bigy}).  It is easy to see that the $C$ term yields a finite and constant limit for $\ze$, but the $D$ term is proportional to $y$ and therefore fades away with time.  More precisely
 \bea &\ze& = -\frac{1}{4}C \approx -\frac{1}{4}A\cos 2q,~{\rm or}~ \nonumber \\
 &\ze& = \fr{k}{4\sqrt{3} Ha}D = -\frac{q}{4}A\sin 2q, \label{zelimit}  \eea with the reminder that $C$ and $D$ are both of order the thermal amplitude, (\ref{CD}).


How does one connect with the radiation era after reheating?  Here $\Phi$ is given by (\ref{Ph_r}) with $A$ and $B$ replaced by $\tilde A$ and $\tilde B$ to distinguish the amplitudes from the pre-inflationary radiation ones, and $y$ and $q$ replaced by new ones defined as \beq \tilde y=\fr{a}{\aeq},~\tilde q=\fr{k}{\sqrt{3} \Heq\aeq}. \label{rdefn} \eeq From (\ref{zedef}), one can proceed to calculate $\ze$, as  \beq \ze = \fr{3\Phi}{2} = \fr{\tilde A}{4} + \fr{\tilde B}{2x} \label{a'} \eeq where $x=\tilde q \tilde y$ is $\ll 1$ when a mode is outside the horizon (note that it is no longer $H$, but rather $Ha^2 = H_{\rm eq} a_{\rm eq}^2$, that remains constant in the radiation era).  Now it has been demonstrated (\citet{mal03}) that for reheating at a constant rate of energy transfer from inflaton to radiation there exists a mode of constant $\ze$.  This mode matches the $C$ mode of (\ref{zelimit}) in both $\ze$ and $\dot\ze$ at the start of reheating (the $D$ mode is already redshifted to an unobservable amplitude by then) and with the $\tilde A$ mode of (\ref{a'}) at the end of reheating provided: \beq \tilde A=-C. \label{match} \eeq  Consequently, for the ensuing radiation era solution one may set \beq \ze = -\fr{C}{4}, \label{Thze} \eeq and hence
\beq \Phi=-\fr{C}{2x^2}\left(\fr{\sin x}{x}-\cos x\right). \eeq  One then finds
 \bea \de &=& -2(x\Phi'+\Phi+x^2\Phi) \nonumber \\
 &=& \fr{2C}{x^3}(-2\sin x+2x\cos x+2x^2\sin x-x^3\cos x), \label{ThPh}
 \eea
where now $\Phi'=\pa\Phi/\pa x$.  Thus we see that the amplitude of the fluctuations after re-entry (when $x=1/\sqrt3$) is approximately the same as it was at horizon exit.   Far ahead of re-entry ($x \ll 1$), however, the density contrast and the velocity divergence  of (\ref{th_L}) are given by \beq \de = \de_r = \fr{C}{3};~\th_r =  -\fr{k^2}{12 \Heq \aeq^2} C = -\fr{\tilde q^2 \Heq}{4} C, \label{Thth} \eeq where in obtaining $\th_r$ use was made of the relation $\th_r = k^2(x\Phi' + \Phi)/(2\Heq \aeq^2)$.

For completeness, it can also be shown that the (unobservable) $\tilde B$ solution of (\ref{a'}) and (\ref{Ph_r}) yields the variables \beq \de = \de_r = \fr{2\tilde B}{x^3}= \fr{2\tilde B}{\tilde q^3 \tilde y^3};~{\rm and}~\th_r = -\fr{3H\tilde B}{2x} = -\fr{3H\tilde B}{2\tilde q \tilde y}. \label{B'} \eeq

\section{The synchronous gauge}

There are three evolution equations in this gauge that can be derived using Newtonian physics and the Copernican Principle (see Appendix A): \bea &\dot\de_r^S& = -\frac{4}{3}\th_r^S -\frac{2}{3}\dot h; \nonumber \\
&\dot\th_r^S& + H\th_r^S = \frac{k^2}{4a^2}\de_r^S; \nonumber \\
&\ddot h& + 2H\dot h = -16\pi G\rho_r \de_r. \label{synch}  \eea  The script $S$ is used to distinguish the radiation density contrast $\de_r^S$ and divergence of the peculiar velocity $\th_r^S=i{\bf k}\cdot {\bf v}_r^S$ from their corresponding values in the longitudinal gauge, $\de_r$ and $\th_r$, which are unscripted. The metric perturbation $\eta$ of (\ref{h}) is fixed once $\de_r^S$, $\th_r^S$, and $\dot h$ are, by the relations \beq \eta = -\fr{4\pi Ga^2\rho\de_S}{k^2} + \fr{Ha^2}{2k^2}\dot h;~\dot\eta=\fr{4\pi Ga^2}{k^2} (\rho + p)\th, \label{eta_defn} \eeq which in this instance reduces to \beq \eta = -\fr{3H^2 a_i^2}{2k^2}\left(\fr{a_i}{a}\right)^2 \de_r^S + \fr{Ha^2}{2k^2}\dot h;~\dot\eta=\fr{2H^2 a_i^2}{k^2}\left(\fr{a_i}{a}\right)^2 \th_r^S. \label{eta}  \eeq
In general, a {\it third} order ordinary differential equation results from the elimination from (\ref{synch}) of two of the three perturbation variables $\de_r^S$, $h$, and $\th_r^S$.

In the radiation era preceding inflation, when the modes are subhorizon, it is usual to assume that \beq \dot\de_r^S \approx -\fr{4}{3}\th_r^S,~{\rm or}~ |\dot\de_r^S| \gg |\dot h|, \label{crit1}  \eeq in which case one can combine (\ref{synch}a) with (\ref{synch}b) to obtain the second order equation (\citet{pee80}) \beq \ddot \de_r^S + H\dot\de_r^S = -\fr{k^2}{3a^2} \de_r^S. \eeq  The solution of (\ref{synch}a) and (\ref{synch}b) is \beq \de_r^S = |\de_r^{\rm th}|e^{i\varphi (t)};~\varphi= \int \fr{k}{\sqrt{3}a}dt + \varphi_0, \label{soln1}  \eeq and $\th_r^S = -3\dot\de_r^S/4$.  To solve for $h$, substitute (\ref{soln1}) into (\ref{synch}c) to convert the latter to \beq \fr{1}{a^2}\fr{d}{dt} (a^2 \dot h) = -\fr{3}{2t^2}e^{i\varphi (t)}, \eeq with the solution \beq \dot h=3|\de_r^{\rm th}|\fr{{\rm Ei} (i\varphi)}{t} + \fr{h_0}{t}, \eeq where Ei$(x)$ is the exponential integral function.

For subhorizon modes with $k \gg Ha$, hence $\varphi \gg 1$, Ei$(i\varphi)$ approaches the limit $i\pi$, so that \beq \dot h = \fr{3i\pi |\de_r^{\rm th}|}{t} \approx 6i \pi H|\de_{\rm th}| + h_0 H. \label{happrox}  \eeq As a self-consistency check, it is easy to verify that (\ref{happrox}) and (\ref{soln1}) satisfy (\ref{crit1}) provided \beq |h_0| \ll |\de_r^{\rm th}| \fr{k}{H_i a_i}. \label{h0}  \eeq  (here the reader is reminded of the fact that $t_i$ marks the end of radiation domination and the onset of inflation).  Unless the universe was born with a large $|h_0|$, however, it is reasonable to assume that \beq |\dot h| \approx H|\de_r^{\rm th}| \label{hlimit}   \eeq during the radiation era between Big Bang and inflation.

At times $t \gg t_i$ when inflation takes over and $H$ is very approximately constant, the right side of (\ref{synch}c) becomes negligibly small.  The solution of (\ref{synch}) is, for $\de_r^S$ and $h$, \beq h=\fr{b}{a^2} + c;~\de_r^S = f\cos\left(\fr{k}{\sqrt{3} Ha} + g\right) - \fr{32 \pi G \rho_v b}{3k^2}. \label{soln2}  \eeq  To maintain continuity with the thermal oscillations of the radiation era of $t \ll t_i$, it is clear that $f \approx |\de_r^{\rm th}|$ and $b$ must be chosen to secure a smooth connection with (\ref{hlimit}) at $t=t_i$, {\it viz.}~\beq |b| \approx a_i^2 |\de_r^{\rm th}|, \label{b}   \eeq  which renders the last term of (\ref{soln2}) negligible.  When $t>t_i$, (\ref{crit1}) continues to hold, because $\de_r^S$ for both independent modes of perturbation oscillate with the amplitude $|\de_r^{\rm th}|$ until horizon exit and becoming either a constant or fading away as $a^{-1}$ afterwards (see below), while $\dot h$ is damped more quickly as $a^{-2}$.

Let us examine more closely the behavior after horizon exit.  The two possible orthogonal solutions from (\ref{soln2}), {\it viz.}~$g=0$ and $g=-\pi/2$ then become, to a high degree of accuracy for $\de_r^S$ and $\th_r^S$, \beq (\de_r^S,~\th_r^S) = \left(1,-\fr{3Hq^2}{4y^2}\right) |\de_r^{\rm th}|=\left( 1,-\fr{k^2}{4H a^2} \right) |\de_r^{\rm th}|, \label{exit1} \eeq and \beq (\de_r^S,\th_r^S) = \left( \fr{q}{y},-\fr{3H}{4}\fr{q}{y} \right) |\de_r^{\rm th}|=\left(\fr{k}{\sqrt{3} Ha},- \fr{\sqrt{3}k}{4a} \right) |\de_r^{\rm th}| \label{exit2} \eeq
for $t \gg t_i$.  By means of (\ref{eta}), (\ref{b}), (\ref{exit1}) and (\ref{exit2}),  one can estimate or calculate the remaining variables, {\it viz.}~\bea |\dot h| \approx \fr{H|\de_r^{\rm th}|}{y^2} =H\left(\fr{a_i}{a}\right)^2 |\de_r^{\rm th}|;~
|\eta| \approx \fr{|\de_r^{\rm th}|}{q^2}  = \fr{3H^2 a_i^2 |\de_r^{\rm th}|}{k^2}, \label{limit1}  \eea {\it i.e.} $|\dot h| \ll H|\de_r^{\rm th}|$ and $|\eta| \ll |\de_r^{\rm th}|$ for both pairs of $(\de_r^S,\th_r^S)$ in (\ref{exit1}) and (\ref{exit2}).  Moreover \beq \dot\eta = -\fr{H}{2y^4} |\de_r^{\rm th}|= -\fr{H}{2}\left(\fr{a_i}{a}\right)^4 |\de_r^{\rm th}|, \label{limit2} \eeq and \beq \dot\eta=-\fr{H}{2qy^3}|\de_r^{\rm th}|=-\fr{\sqrt{3}H}{2}\left(\fr{a_i}{a}\right)^3 \fr{Ha_i}{k} |\de_r^{\rm th}| \label{limit3}  \eeq for the first and second pair respectively.  It is then clear that \beq |\dot\eta| \ll H|\de_r^{\rm th}| \label{limit4}  \eeq for both pairs.  The consequences of (\ref{limit1}) through (\ref{limit4}) will become apparent as we connect the two solutions here with those of the longitudinal gauge.

\section{Proof of gauge consistency}

During the inflation era when a mode is expelled from the horizon, there {\it could} in principle be significant differences in the perturbation variables as compared between the synchronous and longitudinal gauge.  To check for consistency, one appeals to the gauge transformation equations (\citet{ma95}), {\it viz.}~ \beq \de_r^S = \de_r + \fr{2Ha^2(\dot h + 6\dot\eta)}{k^2};~\th_r^S = \th_r - \fr{\dot h +6\dot\eta}{2}, \label{transf1}  \eeq and \beq \ze = \eta - \frac{1}{4}\de_r^S. \label{transf2}  \eeq  Comparing the $\de_r$ part of (\ref{exit1}) and (\ref{exit2}) with (\ref{cd}) one finds that \beq \de_r = \de_r^S,~{\rm and}~\th_r = \th_r^S \label{consist}  \eeq is secured by the simple correspondence \beq C=|\de_r^{\rm th}|~{\rm and}~ D=-|\de_r^{\rm th}|, \label{dc} \eeq  consistent with (\ref{CD}).  This is because the $\dot h$ and $\dot\eta$ contributions to (\ref{transf1}) are insignificant for {\it both} mode solutions ({\it i.e.}~both ($\de_r,\th_r$) pairs), as can be verified with the help of (\ref{limit1}) through (\ref{limit4}).  Moreover, with the same correspondence between $C$, $D$, and $|\de_r^{\rm th}|$, (\ref{transf2}) yields the same result for the curvature perturbation $\ze$ as (\ref{zelimit}), due to the smallness of $\eta$ as given by (\ref{limit1}).

However, the situation is a little different in the radiation era after reheating, as $\eta$ plays a key role in the gauge reconciliation.  Here, the two solutions to (\ref{synch}) involve $\de_S = \de_r^S \sim a^2$ and $a$ are better known (see \eg~the last two of (9.121) of \citet{kol90}.  More precisely the former is \beq  (\de_r^S, \th_r^S, h,\dot h) = Q\left(\fr{2}{3} \tilde y^2, \fr{\Heq}{6} \tilde q^2 \tilde y^2, -\tilde y^2, -2 H_{\rm eq} \right), \label{s1} \eeq where $Q$ is a normalization factor and $\tilde y$ and $\tilde q$ are as defined in (\ref{rdefn}).  Note that the sharp rise of $\th_r^S$ w.r.t. $\tilde q$ (or $k$) explains why (\ref{synch}a) is dominated by the $\dot h$ term on its right side before re-entry, but by the $\th_r^S$ term after.  The variables $\eta$ and $\dot\eta$ of (\ref{eta_defn}) are \beq \eta = -\fr{2}{3\tilde q^2}Q;~\dot\eta = \fr{\Heq}{9}Q. \label{doteta} \eeq  One may now apply (\ref{transf1}) and (\ref{doteta}) to transform $\de_S$ and $\th_r^S$ to the longitudinal gauge.  The results are in agreement with (\ref{Thth}) provided $Q$ is set at \beq Q=\fr{3\aeq^2 \tilde q^2}{8\Heq^2} C. \label{Q} \eeq   Such a choice of $Q$ would also enable one to calculate $\ze$ using the synchronous gauge variables and (\ref{transf2}).  This leads to $\ze=-C/4$, the same as (\ref{Thze}). Moreover, by (\ref{dc}), $\ze$ has the amplitude of the density fluctuations in the thermal radiation on the horizon, at the time of horizon exit of the scale in question.

The second solution $\de_S= \de_r^S \sim a$ is \beq  (\de_r^S, \th_r^S, h,\dot h) = P\left(\fr{8\tilde y}{9}, 2\tilde y H, -\fr{16}{3}\tilde y, -\fr{16}{3} \tilde y H\right), \label{s2} \eeq  for some $P$.  Consistency with the longitudinal gauge solutions for $\de_r$, $\th_r$, and $\ze$,~{\it viz.} (\ref{B'}) and the $\tilde B$ term of (\ref{a'}), is secured by the transformations equations of (\ref{transf1}) and (\ref{transf2}), after setting $P=-3\tilde B \tilde q/8$.

To the best of the author's knowledge this is the first time the longitudinal and synchronous gauge are formally reconciled in detail by explicit gauge transformation.  Previous efforts include \citet{ma95} who demonstrated consistency of results upon horizon re-entry by numerical technique, and \citet{wei03} who was only concerned with the adiabatic (constant) mode at times when it is outside the horizon.

\section{Discussion: criterion for the start of standard inflation}

The key message of the past 3 sections is the conservation of the longitudinal gauge density contrast $\de_r$ of relic radiation, and of $\ze$, for one adiabatic mode of perturbation when the scale of interest is outside the horizon, {\it i.e.}~(\ref{Thth}), (\ref{dc}), and (\ref{zedef}).  Thus, if a pre-inflationary radiative FRW stage existed, there will be O(1) fluctuations on scales comparable with the peak wavelength of the black body spectrum, {\it i.e.}~$k\sim \Ti$.  By (\ref{Ti}) one gets
 \beq \frac{1}{k} \approx 3e^N/T_0. \eeq
For $N=60$ this scale would be a few Mpc.   For more e-folds of inflation the observed cosmological scales would have originated from {\it sub}-wavelength ones, $k > T$, on which the density contrasts are {\it above} unity, see \citet{lie13}.   Even restricting oneself to the classical regime of $N \leq 60$, {\it viz.}~the scope of this paper, it is still the case that the pre-existence of radiation domination means the {\it total} density contrast at the onset of inflation $t_i$ is of order unity or larger on the relevant scales, and they would invalidate linear growth equations like (\ref{evol0}) and (\ref{evol1}), leaving one with no quantitative way of evolving the perturbations from $t_i$ onwards.  At best, one could only heuristically appeal to causal interactions within the radiation fluid as a mechanism that might have maintained the statistical fluctuation amplitudes of the radiation density at their thermal values ahead of the horizon exit of the corresponding scales.  But without the support of any mathematically rigorous formalism this argument is far from being secure.  It tells us nothing about how {\it e.g.} the scalar and other modes of the two-fluid medium evolve before each mode exits (see below for more elaboration), and then there is the question of what happens after exit.  The other `way out' is to postulate $N \ll 60$, {\it i.e.}~the scale of the GUT horizon is inflated by just the right amount to match today's horizon.  Such a contrivance would reduce the usefulness of the inflationary hypothesis to the same level as the Anthropic Principle.

It might be thought sufficient to reinstate linear theory that the gravitational potential $\Phi\approx Ga^2\de\rho/k^2 \ll 1$.  Beware however that the sound speed $c_s^2 = \de P/\de\rho$ is an indispensable coefficient of (\ref{evol0}), {\it i.e.}~ without precise knowledge of $c_s^2$ (\ref{evol0}) is useless; yet the classical value of $c_s^2$ becomes very questionable on sub-wavelength scales when $\de = \de\rho/\rho \geq 1$ and likewise for $P$.  A similar argument can be made about the growth equations in the synchronous gauge, \eg~(\ref{synch}) acquires an extra {\it non-linear} term of the form $\dot h \de_r^S$ that {\it must} be included when $\de_r^S \geq 1$.  The situation is worse than this.  When the {\it relative} density contrast $\de_i$ of each fluid component is of order 1 or larger, then despite $\Phi \ll 1$ the vector and tensor modes are coupled to the scalar modes and become important (Bertschinger 1993, Mollerach and Matarrese 1997).

To further expose the inadequacy of the standard classical treatment, one should fully expect very significant fluctuations on the sub-wavelength scales, as indicated by the change in form of the power spectrum from (\ref{PSmallk})  to (\ref{DeLargek}) on the scale of a wavelength, {\it viz.} $\cP(k) \sim k^4$ for $k\gg T_i$, {\it i.e.}~$\de_r > 1$ on these very small scales. In this case, one may also think in terms of patches of size $1/k < 1/T$ ($k > T$) inflating at very different start times due to the large variation in the density ratio of radiation to inflaton, resulting in a highly inhomogeneous observable universe.  Although it is not inconceivable that the large thermal effect and the nonlinearity it causes might fortuitously `cancel' to smooth out the perturbations at the end of an exact (non-perturbative) treatment, it seems far more likely that they would be observationally unacceptable.  At least such a `wishful' outcome needs to be demonstrated, but it has not.  For more e-folds than $N \approx 60$, the thermal wavelength scale would be inflated beyond today's Hubble radius.  In that case all cosmological scales would be in the  $k\gg \Ti$ domain.  Moreover, for $N \gg 70$ the relevant scales at the beginning of inflation were sub-Planck, rendering the estimation of the amplitude of fluctuations at that time even more problematic.  To postulate an FRW phase in a sub-Planck-scale universe seems nonsensical.

As indicated in section 2, the foregoing developments were based upon the premise of a thermalized fluid whose large statistical fluctuations  on the initially small scales prevents inflation from having a proper start.  How sensitive is the `false start' conclusion to thermalization?  One may tackle this problem by asking for the power spectrum $\cP(k)$ that marginally reinstates the paradigm.  Evidently this cannot be one that diverges as $\cP(k) \sim k^4$, or equivalently $P(k) \sim k$, see (\ref{PSmallk}).  In fact, the `minimal' $P(k)$ that satisfies the requirement of a finite  $\xi ({\bf r})$ at small scales, with the two-point function $\xi ({\bf r})$ defined as the Fourier counterpart of $P({\bf k})$, is evidently $P({\bf k}) \sim 1/k^3$ or $\cP ({\bf k}) =$~constant.  But since the value of this constant is determined to be of order unity by matching against the O(1) fluctuations on the transition scale of $k \sim \Ti$, to revalidate linear theory one must invoke $P(k) \sim 1/k^n$ with $n>3$, or more precisely  \beq \cP({\bf k}) \ap \left(\fr{T}{k}\right)^{n-3}~{\rm with}~n>3,~{\rm for}~k\gg T, \label{start} \eeq to replace (\ref{DeLargek}), which will enable $\de_r$ to become $\ll 1$ on scales $k \gg \Ti$.  Such a super-homogeneous pre-inflationary state is a far-cry from thermal, but has been considered by \cite{gio12}.

\section{Conclusion}

A comprehensive calculation of the evolution of pre-inflationary perturbations from subhorizon to superhorizon scales in the longitudinal and synchronous gauge is presented, with gauge reconciliation at a level of detail not easily accessible in the published literature (additionally an Appendix is also provided to show how the growth equations in the synchronous gauge can be derived without appealing to the Einstein Field Equations). It is concluded that the standard `minimal' version of inflation and reheating cannot involve an earlier radiation dominated FRW phase.

There are obvious ways in which the theory could avoid these problems.  Most simply, the universe could be born inflating (\citet{vil82,har83}).  Later versions of inflation, such as chaotic inflation (\citet{lin83,lin86}) do not necessarily begin with a pre-inflationary FRW phase, and would therefore be unaffected.  Yet the point is that, as explained in the introduction, if the smoothest possible component of massless particles already caused such severe problems, it is hard to imagine how the presence of {\it any} equipartition field of other matter-like particles, be they curvaton or relics, would be more accommodating to the theory.   Thus the claim that inflation also simultaneously solved such other major problems of cosmology as flatness and the absence of monopoles is overstated.

In a variant scenario that reinstates the {\it status quo}, one could invoke a super-homogeneous state of power spectrum $P(k) \sim 1/k^n$ with $n > 3$.  Although \citet{gio12} argued the plausibility of such a state, postulating its existence would add further contrivance to an already long list of initial conditions necessary for `successful' inflation.  In this instance, if the pre-inflationary fluid is already super-homogeneous the theory again cannot be deemed to have `explained' the flatness problem and the absence of relics.  

\section*{Acknowledgments}

I wish to acknowledge helpful correspondence with Tom Kibble, Massimo Giovannini, Michael Joyce, Joao Magueijo, and David Wands, as well as critical comments from Andrei Linde and David Lyth.

\appendix

\section{Newtonian derivation of the growth equations synchronous gauge}

We restrict our treatment to a radiation dominated universe.  The continuity equation of mass and momentum conservation is \beq \fr{\partial\rho}{\partial t} + {\bf \nabla}\cdot [(\rho+p) {\bf v}] = 0, \label{A1} \eeq where $\nabla = \nabla_{\bf R} = \pa/\pa {\bf R}$ with \beq {\bf R} = a{\bf r} \label{R} \eeq as the physical distance and $\pa/\pa t$ is done with ${\bf R}$ held fixed, {\it i.e.}~ \beq \fr{\pa\rho}{\pa t}\bigg|_{\bf R} = \fr{\pa\rho}{\pa t}\bigg|_{{\bf r}} + \dot {\bf r} \cdot \fr{\pa\rho}{\pa {\bf r}} = \fr{\pa\rho}{\pa t}\bigg|_{\bf r} - H {\bf r}\cdot\nabla_{\bf r}~\rho,  \eeq where use was made of $\nabla_{\bf R} = \nabla_{\bf r}/a$, and $\dot{\bf r}|_{\bf R} = -\dot a {\bf R}/a^2 = -H {\bf r}$.   Hence (\ref{A1}) becomes \bea \fr{\pa\rho}{\pa t}\bigg|_{\bf r} &-& H {\bf r} \cdot\nabla_{\bf r} ~\rho + \fr{\rho+p}{a} \nabla_{\bf r} \cdot (Ha {\bf r} + {\bf u})  \nonumber\\
&&\ + \fr{Ha {\bf r}{\bf u}}{a}\cdot\nabla_{\bf r} (\rho+p) = 0, \eea where we wrote \beq {\bf v} = \dot{\bf R} = Ha {\bf r} + {\bf u} \label{v} \eeq with ${\bf u} = a\dot {\bf r}$ being the peculiar velocity.

It is now necessary to discuss the diverging ${\bf r} \cdot\nabla_{\bf r}~p$ term, which is to do with the fact that in terms of the
cosmic time t and the physical coordinate ${\bf R} = a {\bf r}$ the metric is $ds^2 = dt^2 - (dR - HRdt)^2$, and has cross terms, {\it i.e.}~$g_{00} = 1 - H^2R^2$ and $g_{0j} = HR^j$.  That means there should be extra terms in the energy and momentum conservation equations.  Fortunately, near any
particular location, matter should behave more or less as it does in flat
 space.  In terms of local coordinates t and ${\bf R}$ the equations will be the
 same as flat space-time for small R.  There will be differences for
 large R, but if one is interested in equations connecting the local
 quantities here, one can legitimately set ${\bf R} = 0$ (or equivalently ${\bf r}=0$) and does not
need to work out the equations for other values of ${\bf R}$ separately,
 because the cosmological principle tells us that in terms of their locally
 defined variables, the equations would be the same as they are at the origin.

 Undertaking this step, and dropping henceforth the suffix ${\bf r}$ from $\pa\rho/\pa t$ and $\nabla$, we obtain
 \beq \fr{\pa\rho}{\pa t} + 3H(\rho + p) +\fr{\rho+p}{a}\nabla_{\bf r} \cdot {\bf u} +\fr{{\bf u}}{a}\cdot\nabla_{\bf r}~p = 0.  \eeq  In a homogeneous background universe the $\nabla$ terms vanish, and so to lowest order one recovers the familiar equation $\pa\rho/\pa t = -3H(\rho + p)$.  The next order is, for a radiation dominated universe with $\rho + p = 4\rho/3$, \beq \de\dot\rho = -4\rho\de H - 4H\rho\de - \fr{4i k_j u^j}{3a}\rho =-\frac{2}{3}\rho \dot h -4H\rho\de-\frac{4}{3}\rho\th, \label{almost} \eeq where \beq \th = i\fr{{\bf k}\cdot {\bf u}}{a} \label{expansion} \eeq is the velocity divergence (expansion).  (\ref{almost}) can be compressed to read \beq \dot\de = -\frac{2}{3}\dot h -\frac{4}{3} \th, \label{1st} \eeq which is the same as the first of (\ref{synch}).  In general, for a medium with any equation of state, (\ref{1st}) becomes \beq \dot\de + 3(c_s^2 -w)H\de = -(1+w)\left(\th + \fr{\dot h}{2}\right), \eeq where $c_s^2 = dP/d\rho$ is the speed of sound, and $w = p/\rho$.

 To get the remaining equations of (\ref{synch}) we turn to the Euler equation of momentum flow: \beq \fr{\pa {\bf v}}{\pa t}\bigg|_{\bf R} + ({\bf v}\cdot \nabla_{\bf R}){\bf v} = -\fr{1}{\rho+p} \nabla_{\bf R}~p - \nabla_{\bf R} \Phi, \label{Euler} \eeq where $\Phi$ is the {\it total} gravitational potential here.  Bearing in mind again (\ref{R}) and (\ref{v}), we repeat similar exercises as before: \bea \fr{\pa {\bf v}}{\pa t}\bigg|_{\bf R} &=& \fr{\pa {\bf v}}{\pa t}\bigg|_{\bf r} - H( {\bf r}\cdot\nabla_{\bf r}) {\bf v};~({\bf v} \cdot\nabla_{\bf R}){\bf v} \nonumber \\
 &=& H({\bf r} \cdot\nabla_{\bf r}){\bf v} +
 \left(\fr{{\bf u} \cdot\nabla_{\bf r}}{a}\right) {\bf v}. \eea  Dropping once more the suffix ${\bf r}$ from $\pa\rho/\pa t$ and $\nabla$, \beq \fr{\pa {\bf v}}{\pa t} + \left(\fr{{\bf u} \cdot\nabla}{a} \right) {\bf v} = -\fr{1}{\rho+p} \nabla p - \fr{1}{a}\nabla \Phi. \label{Euler1} \eeq Next, enlist (\ref{v}) another time to write \beq \fr{\pa {\bf v}}{\pa t} = \ddot a {\bf r} +\dot{\bf u} + H {\bf u}; \label{dvdt} \eeq and moreover  \beq \fr{{\bf u} \cdot\nabla}{a} {\bf v}  = \fr{{\bf u} \cdot\nabla}{a} (Ha {\bf r}) + \fr{{\bf u} \cdot\nabla}{a} {\bf u}. \eeq  The last term is second order of small quantities.  The $Ha {\bf r}$ term gives \bea \fr{{\bf u} \cdot\nabla}{a}  {\bf v} &=& \fr{{\bf u} \cdot\nabla}{a} (Ha {\bf r}) \nonumber\\
 &=& \fr{H}{a}\left(a \dot x \fr{\pa}{\pa x} + a \dot y \fr{\pa}{\pa y} + a \dot z \fr{\pa}{\pa z}\right) (ax {\bf i} + ay {\bf j} + az{\bf k}) \nonumber \\
 &=& Ha \dot {\bf r} \nonumber\\
 &=& H {\bf u}. \label{Hu} \eea (\ref{Euler1}) now reads \beq \ddot a  {\bf r} + \dot{\bf u} + 2H {\bf u} =
-\fr{1}{\rho+p} \nabla p - \nabla \Phi. \eeq   The background (lowest order) terms give the usual deceleration parameter $q$, with $\nabla\Phi$ interpreted as a force that depends on the nature of the cosmic substratum.  The first order term is \beq \dot{\bf u} + 2H {\bf u} = -\fr{1}{a(\rho+ p)} \nabla (\de p);~{\rm or}~ \dot\th + H\th = \fr{k^2}{4a^2} \de, \eeq where the last equation is the same as (\ref{synch}b) for any radiation era with $p/\rho = dP/d\rho = 1/3$, and is obtained by means of (\ref{expansion}) and the synchronous gauge condition (sometimes called `free fall' condition) $\th = k^2 \de\Phi/(2Ha^2)$.  For a general medium this equation is \beq \dot\th + (2-3w)H\th = \fr{k^2 c_s^2 \de}{(1+w) a^2}. \eeq

Finally, (\ref{synch}c) is just the perturbed version of the Friedmann equation, which for the radiation era is \beq \dot H + H^2 = -\fr{8\pi G}{3} \rho. \eeq  Specifically by defining $\de H = \dot h/6$  and taking account of the rate equation $\dot\rho = -4H\rho$, one obtains (\ref{synch}c), the general form of which is \beq \ddot h + 2H\dot h = -3H^2(1+3c_s^2)\de. \eeq

\end{document}